\documentstyle[12pt]{article} 

\begin{document}

\baselineskip=20pt plus 1pt minus 1pt 
\textwidth=6.5truein
\textheight=9.0truein

\centerline{\bf GENERALIZED DEFORMED OSCILLATOR }
\centerline{\bf FOR VORTICES IN SUPERFLUID FILMS} 

\bigskip\bigskip
\centerline{DENNIS BONATSOS \footnote[1]{Also at the European 
Centre for Theoretical Studies in Nuclear Physics and Related Areas 
(ECT$^*$), Strada delle Tabarelle 286, I-38050 Villazzano (Trento), Italy.
E-mail: bonat@ectstar.ect.unitn.it, bonat@cyclades.nrcps.ariadne-t.gr}}

\centerline{ Institute of Nuclear Physics, NCSR ``Demokritos''}

\centerline{GR-15310 Aghia Paraskevi, Attiki, Greece}

\bigskip
\centerline{C. DASKALOYANNIS \footnote[2]{E-mail: 
daskaloyanni@olymp.ccf.auth.gr} }

\centerline{ Department of Physics, Aristotle University of 
Thessaloniki}

\centerline{GR-54006 Thessaloniki, Greece}

\bigskip\bigskip\bigskip

\centerline{\bf Abstract}

The algebra of observables of a system of two identical vortices in a
superfluid thin film is described as a generalized deformed oscillator 
with a structure function containing a linear (harmonic oscillator) term 
and a quadratic term. In contrast to the deformed oscillators occuring in 
other physical systems (correlated fermion pairs in a single-$j$ nuclear shell,
Morse oscillator), this oscillator is not amenable to perturbative treatment 
and cannot be approximated  by quons. From the mathematical viewpoint, this 
oscillator provides a novel boson realization of the algebra su(1,1).  

\bigskip\bigskip

% PACS numbers: 03.40Gc, 67.40Vs, 02.20Sv

\vfill\eject

The possibility of the  existence in two dimensions of statistics intermediate 
between the bosonic and fermionic ones has attracted recently much attention
$^1$,
since it is related to questions concerning the fractional quantum Hall 
effect and anyon superconductivity. The quantization of the system of 
two identical vortices has also attracted much attention, since it has been
suggested as an anyon candidate $^{2-7}$, although 
opposing views also exist $^8$. 

On the other hand, quantum algebras (quantum groups) $^{9,10}$
 which are nonlinear
generalizations of the usual Lie algebras to which they reduce for 
appropriate values of the deformation parameter(s), have been finding 
applications as the dynamical symmetry algebras of several physical 
systems. For the boson realization of these algebras, various kinds 
of deformed oscillators have been introduced $^{11-15}$
and unification schemes for them have been suggested (see $^{16}$ for 
a list of references). In the framework of 
intermediate statistics the term quons has been used for some of 
these oscillators $^{17,18}$. 

In the present work the quantization of a system of two identical vortices 
in a superfluid thin film will be considered. It will be demonstrated that 
the algebra of the observables of this system is a generalized deformed 
oscillator $^{16,19}$ with a structure function containing a linear 
term, which corresponds to the usual harmonic oscillator, and a quadratic 
term.  
This oscillator is not amenable to perturbative treatment and cannot be 
approximated by quons, which  do provide an adequate description of other
physical systems. 

Let us consider a system of two identical massless point vortices in an 
unbounded flat, frictionless, incompressible, superfluid thin film $^{2,7}$.
If the vortices are centered at the points $(x_i, y_i)$,
$i=1$, 2, we introduce the relative coordinates 
$$ x=x_1-x_2, \qquad y=y_1-y_2,\eqno(1)$$
as well as the center of vorticity coordinates 
$$ X={x_1+x_2\over 2}, \qquad Y={y_1+y_2\over 2}.\eqno(2)$$
The center of vorticity coordinates are constants of motion, while 
the equations of motion for the relative coordinates are
$$ \kappa {dx\over dt} = 2 {\partial H\over \partial y}, \qquad 
   \kappa {dy\over dt} =-2 {\partial H\over \partial x},\eqno(3)$$
where $\kappa$ is the quantized vorticity of the superfluid $^7$, 
given by 
$$\kappa = {2\pi \hbar \over m},\eqno(4)$$
where $m$ is the mass of the atoms in the superfluid. We will assume that 
$\kappa$ is positive. (Negative $\kappa$ corresponds to vortices with 
opposite circulation.) 
The relevant Hamiltonian is $^6$
$$ H= -{\kappa^2 \over 4 \pi} \ln\left({x^2+y^2\over a^2}\right),\eqno(5)$$
where $a$ is a scale parameter. The energy $E$ of the system is related 
to the Hamiltonian $H$ by 
$$ E= \rho \delta H,\eqno(6)$$
where $\rho$ is the density and $\delta$ the thickness of the superfluid 
film. The momentum conjugate to $x$, with appropriate dimensions, is 
$$ p_x = {\kappa \rho\delta \over 2} y,\eqno(7)$$
the canonical quantization condition being 
$$ [x, p_x ]= i\hbar,\quad {\rm or}\quad [x,y]= {2 i\hbar \over \kappa \rho
\delta}.\eqno(8)$$
The relative coordinates $x$ and $y$ are not invariant under the 
interchange of the position of the two vortices 
$$ (x,y) \to (-x, -y), \eqno(9)$$
while the observables of the system have to be invariant under this 
interchange. Therefore quantities acceptable as observables are $^6$
$$ A= {1\over 8} \kappa \rho\delta (x^2+y^2), \eqno(10a) $$
$$ B= {1\over 8} \kappa \rho \delta (y^2-x^2),\eqno(10b) $$
$$ C= {1\over 8} \kappa \rho \delta (xy+yx),\eqno(10c) $$
satisfying the commutation relations 
$$ [A,B]= i\hbar C, \qquad [A,C]=-i\hbar B, \qquad [B,C]=-i\hbar A.\eqno(11) $$
Introducing the operators 
$$ B_{\pm}= B\pm iC,\eqno(12)$$
the commutation relations take the form $^6$ 
$$ [A, B_{\pm} ]= \pm \hbar B_{\pm}, \qquad [B_+, B_-]= -2\hbar A.\eqno(13)$$
Defining 
$$ \tilde A ={A\over \hbar}, \qquad \tilde B_{\pm} = \pm {B_{\pm} \over 
\hbar},\eqno(14)$$
one sees that these commutation relations acquire the su(1,1) form
$$ [\tilde A , \tilde B_{\pm}]= \pm \tilde B_{\pm}, \qquad 
[\tilde B_+, \tilde B_-]= -2 \tilde A.\eqno(15)$$
For later use we also remark that the first of these equations can be written 
as 
$$ [{A\over \hbar}, B_{\pm}] = \pm B_{\pm}.\eqno(16) $$

On the other hand, it is known that a deformed oscillator $^{16,19}$
can be defined by the algebra generated by the
operators $\big\{ 1,a,a^\dagger,N\big\}$ and the {\it structure
function} $\Phi (x)$, satisfying the relations:
$$\left[ a , N \right] = a, \quad  \quad
 \left[ a^\dagger , N \right] = -a^\dagger , \eqno(17)$$
and
$$ a^\dagger a=\Phi(N)=[N], \qquad aa^\dagger=\Phi(N+1)=[N+1], \eqno(18)$$
where $\Phi(x)$ is a positive analytic function with
$\Phi(0)=0$ and $N$ is the number operator.
From Eq. (18)
 we conclude that:
$$N= \Phi^{-1}\left( a^\dagger a \right), \eqno(19)$$
and that the following commutation and
anticommutation relations are obviously satisfied:
$$ \left[ a,a^\dagger \right] = [N+1]-[N], \qquad
   \left\{ a,a^\dagger \right\} = [N+1]+[N] .\eqno(20)$$

The {\it structure  function} $\Phi(x)$ is characteristic to the deformation
scheme.  In Table 1 the structure functions corresponding
to various deformed oscillators are given.

Generalized deformed algebras possess a Fock space of 
 eigenvectors 
$|0>$, $|1>$, $\ldots$, $|n>$, $\ldots$
of the number operator $N$
$$N|n>=n|n>,\quad <n|m>=\delta_{nm}, \eqno(21) $$
if the {\it vacuum state} $|0>$ satisfies the following relation:
$$ a|0>=0. \eqno(22)$$
 These eigenvectors are generated 
 by the formula:
 $$ \vert n >= {1 \over \sqrt{ [n]!}} {\left( a^\dagger \right)}^n \vert 0 >,
\eqno(23) $$
where
 $$[n]!=\prod_{k=1}^n [k]= \prod_{k=1}^n \Phi(k). \eqno(24) $$
The generators  $a^\dagger$ and $a$ are the creation and
destruction operators of this deformed oscillator algebra:
$$a\vert n> = \sqrt{[n]} \vert n-1>,\qquad
 a^\dagger\vert n> = \sqrt{[n+1]} \vert n+1>. \eqno(25) $$

Returning to the system of two vortices, we see that the first  equation in
Eq. (15) 
has the same form as Eq. (17), with $B_+$ ($B_-$) playing the role of the 
creation
(annihilation) operator, and $A$ being related to the number operator
$$ B_+= a^{\dag}, \qquad B_-=a, \qquad {A\over \hbar} = N+u,\eqno(26)$$
where $u$ is a constant to be determined later. 

Performing a straightforward calculation one can show that 
$$ B_+ B_-= A^2 -\hbar A +{3\over 16} \hbar^2.\eqno(27)$$
Substituting in Eq. (27) $A$ from  Eq. (26) one obtains 
$$ B_+ B_-= \Phi(N), \qquad B_- B_+ =\Phi(N+1),\eqno(28)$$
where $\Phi(x)$ is the structure function given by
$$ \Phi(x)= \hbar^2 \left( (x+u)^2 -(x+u)+{3\over 16}\right).\eqno(29)$$
As already mentioned, the structure function has to satisfy the 
conditions 
$$ \Phi(0)=0,\qquad \Phi(x)\geq 0 \quad {\rm for}\quad  x\geq 0,\eqno(30)$$
which imply 
$$u={3\over 4},\eqno(31)$$
and therefore $$\Phi(x)= \hbar^2 \left(x^2+{x\over 2}\right),\eqno(32)$$
while from Eq. (26) one has
$$ A=\hbar \left(N+{3\over 4}\right), \qquad N=0,1,2,\ldots .\eqno(33)$$ 
The basis in the present case is given by 
$$ |n> = {1 \over \sqrt{[n]!}} (B_+)^n |0>, \eqno(34) $$
where $$ [n]! = \prod_{k=1}^n \Phi(k).\eqno(35) $$
In this basis one has 
$$ N |n> = n|n>, \eqno(36) $$
the eigenvalues of  the operator $A$ being
$$ A_n = \hbar \left(n+{3\over 4}\right).\eqno(37)$$ 
Since the Hamiltonian (Eq. 5) can be written as
$$ H = -{\kappa^2 \over 4\pi} \ln \left( {8\over \kappa \rho\delta a^2} 
A\right), \eqno(38)$$
the energy eigenvalues are 
$$ E_n = -{\rho\delta \kappa^2 \over 4 \pi} \ln \left({8\hbar \over \kappa \rho
\delta a^2} \left(n+{3\over 4}\right) \right),\eqno(39)$$
in agreement with Eq. (17) of $^6$.  

We have therefore constructed a deformed oscillator describing the system of 
two vortices. The following comments are now in place:

i) As seen in Table 1, the structure function for usual bosons 
(harmonic oscillator) 
is $\Phi(N)=N$, while in the present case of the system of two vortices one 
has $\Phi(N) = \hbar^2 (N/2+N^2)$, i.e. the structure 
function is a linear combination
of a  bosonic and a quadratic term,  indicating the relation of the 
two-vortex system to intermediate statistics $^{1-5}$.

ii) The algebra of observables studied here is the same as the one occuring 
for a system of two anyons in a strong magnetic field $^{29}$. 

iii) The assumption that vortices are point-like is known to be quite
restrictive $^{28}$. 

iv) Deformed oscillators have been found useful in describing some other 
physical systems. In the case of correlated fermion pairs of zero angular
momentum in a single-$j$ nuclear shell the structure function of the 
relevant oscillator is $^{30}$ 
$$ \Phi(N) =  \left(1+{1\over \Omega}\right)N -{N^2\over \Omega},\eqno(40)$$
where $2\Omega = 2j+1$ is the size of the shell,
while in the case of the Morse oscillator the structure function 
has the form $^{31}$
$$ \Phi(N) = N-x_e N^2,\eqno(41)$$
where $x_e$ is the anharmonicity constant. Both cases can be viewed as 
perturbed harmonic oscillators, since the parameters $1/\Omega$ and 
$x_e$ are small. In the case of the two-vortex system, however, no 
small parameter appears, the $N^2$ term having a coefficient twice as 
large as the one of $N$. Therefore no perturbative treatment is possible 
in this case. 

v) The deformed bosons introduced in the two examples mentioned above
(single-$j$ shell, Morse potential), can be approximated by $Q$-bosons
$^{11,12}$, characterized by the structure function
$$ F(N)= {Q^N-1\over Q-1}, \eqno(42)$$
where $Q=e^T$ (with $T$ real). In the former case the deformation 
parameter $T$ is related to 
the size of the shell by $T=-2/\Omega$ $^{32}$, while in the  latter
it is connected to the anharmonicity constant by $T=-2 x_e$ $^{31}$. In the
present case of the two-vortex system the deformed bosons cannot be 
approximated by $Q$-bosons, due to the lack of a small parameter. $Q$-bosons
are also appearing in the literature of intermediate statistics under the name
 of quons $^{17,18}$. 

vi) It should always be kept in mind that in the cases of the two examples
mentioned above the energy of the system is given by the energy of the 
oscillator, while in the case of the two vortices the energy of the 
system is related to the number operator of the deformed oscillator. 

vii) The deformed oscillator operators $B_+$, $B_-$ and $N$ provide a new 
boson mapping for the generators of the su(1,1) algebra of Eq. (15), through 
$$ \tilde B_+ = {a^{\dag}\over \hbar},\qquad \tilde B_-={a\over \hbar},  
\qquad \tilde A = N+{3\over 4}.\eqno(43)$$
Existing boson mappings $^{33-35}$ of su(1,1) in terms of usual bosons include 
the Schwinger realization 
$$ \tilde B_+ = a_1^{\dag} a_2^{\dag}, \qquad \tilde B_-=a_1 a_2, \qquad 
\tilde A = {1\over 2} (N_1+N_2+1), \eqno(44)$$
where
$$ N_1= a_1^{\dag} a_1, \qquad N_2=  a_2^{\dag} a_2,\eqno(45)$$
which involves two sets of bosons, ($a_1^{\dag}$, $a_1$, $N_1$ and 
$a_2^{\dag}$, $a_2$, $N_2$), as well as the realization 
$$ \tilde B_+= {1\over 2} (a^{\dag})^2, \qquad \tilde B_- = {1\over 2} 
a^2, \qquad \tilde A = {1\over 2} \left(N+{1\over 2}\right),\eqno(46)$$
with $$N=a^{\dag} a, \eqno(47)$$ 
involving one boson. In both of these realizations the su(1,1) operators 
are mapped onto bilinear combinations of bosons, while in the present case 
they are mapped onto terms linear in  the deformed bosons.   
 
viii) An expression similar to Eq. (39) for the energy of a pair of interacting
massless vortices in two dimensions has been derived in $^{36}$, where 
different quantization schemes have been compared. This energy is given by
$$ E_N =-{1\over 4} \quad q^2 \quad {n_1 n_2 \over 2\pi} \quad 
\ln\left( {2c\over r_0^2}
\quad (2N+1)\right), \eqno(48)$$
where $q^2=\rho_0 \sigma^2$, $\rho_0$ is the fluid mass-density, $\sigma
=h/m$ is the vorticity (same as $\kappa$ of Eq. (4) in our notation), 
$r_0$ is the vortex core size, and $n_i$ are integers describing the 
vorticity quantization $^{2,36,37}$. 

ix) A canonical quantization scheme for vortices in superfluid He II 
has been given in $^{38,39}$. An {\sl infinite} Lie algebra of incompressible 
flows is obtained, introduced by the quantization in the theory.  

In conclusion, we have shown that the algebra of observables of a system of 
two identical vortices in a superfluid thin film can be described in terms 
of a generalized deformed oscillator characterized by a structure function
containing a linear (harmonic oscillator) term and a quadratic term.  
From the mathematical point of view, this deformed
oscillator provides a novel boson realization of the su(1,1) algebra. 
In contrast to other physical systems (correlated fermion pairs, Morse 
oscillator) the two-vortex system cannot be seen as a perturbed harmonic 
oscillator and cannot be approximated by $Q$-bosons or quons. 

{\bf Acknowledgment}

Support by the Greek Secretariat of Research and Technology under contract 
PENED95/1981 and  by CEC under contract ERBCHBGCT930467 (DB)
is gratefully acknowledged.

\vfill\eject

\centerline{\bf References}

\def\ref{\par\hangindent=1.0cm\hangafter=1}
\parindent=0pt

\ref {1.} 
J. M. Leinaas, {\it Phys. Reports} {\bf 242}, 371 (1994). 

\ref {2.}
R. Y. Chiao, A. Hansen and A. A. Moulthrop, {\it Phys. Rev. Lett.} {\bf 54}, 
1339 (1985). 

\ref {3.}
A. Hansen, R. Y. Chiao and A. A. Moulthrop, {\it Phys. Rev. Lett.} {\bf 55}, 
1431 (1985). 

\ref {4.}
F. D. M. Haldane and Y. S. Wu, {\it Phys. Rev. Lett.} {\bf 55}, 2887 (1985). 

\ref {5.}
G. A. Goldin, R. Menikoff and D. H. Sharp,  {\it Phys. Rev. Lett.} {\bf 58}, 
174 (1987).  

\ref {6.}
J. M. Leinaas and J. Myrheim, {\it Phys. Rev. B} {\bf 37}, 9286 (1988). 

\ref {7.}
L. Onsager, {\it Nuovo Cimento} {\bf 6}, Suppl., 279 (1979). 

\ref {8.}
J. Dziarmaga, cond-mat/9504004. 

\ref {9.}
V. G. Drinfeld, in {\it Proceedings of the International Congress of 
Mathematicians}, ed. A. M. Gleason (American Mathematical Society,
Providence, RI, 1986), p. 798. 

\ref {10.}
M. Jimbo, {\it Lett. Math. Phys.} {\bf 11}, 247 (1986). 

\ref {11.}
M. Arik and D. D. Coon, {\it J. Math. Phys.} {\bf 17}, 524 (1976). 

\ref {12.}
V. V. Kuryshkin, {\it Annales de la Fondation Louis de Broglie} {\bf 5}, 111 
(1980). 

\ref {13.}
L. C. Biedenharn, {\it J. Phys. A} {\bf 22}, L873 (1989). 

\ref {14.}
A. J. Macfarlane, {\it J. Phys. A} {\bf 22}, 4581 (1989). 

\ref {15.}
C. P. Sun and H. C. Fu, {\it J. Phys. A} {\bf 22}, L983 (1989). 

\ref {16.}
D. Bonatsos and C. Daskaloyannis, {\it Phys. Lett. B} {\bf 307}, 100 (1993). 

\ref {17.}
O. W. Greenberg, U. Maryland preprint 93-097 (1993). 

\ref {18.}
S. Meljanac and A. Perica, {\it J. Phys. A} {\bf 27}, 4737 (1994). 

\ref {19.}
C. Daskaloyannis, {\it J. Phys. A} {\bf 24}, L789 (1991). 

\ref {20.}
R. Chakrabarti and R. Jagannathan, {\it J. Phys. A} {\bf 24}, L711 (1991). 

\ref {21.}
Y. Ohnuki and S. Kamefuchi, {\it Quantum Field Theory and Parastatistics}
(Springer-Verlag, Berlin, 1982). 

\ref {22.}
K. Odaka, T. Kishi and S. Kamefuchi, {\it J. Phys. A} {\bf 24}, L591 (1991). 

\ref {23.}
A. Jannussis, G. Brodimas, D. Sourlas and V. Zisis, {\it Lett. Nuovo Cimento}
{\bf 30}, 123 (1981). 

\ref {24.}
T. Hayashi, {\it Commun. Math. Phys.} {\bf 127}, 129 (1990). 

\ref {25.}
D. Bonatsos and C. Daskaloyannis, {\it J. Phys. A} {\bf 26}, 1589 (1993). 

\ref {26.}
K. S. Viswanathan, R. Parthasarathy and R. Jagannathan, {\it J. Phys. A}
 {\bf 25}, L335 (1992). 

\ref {27.}
R. Chakrabarti and R. Jagannathan, {\it J. Phys. A} {\bf 27}, L277 (1994). 

\ref {28.}
G. A. Goldin, R. Menikoff and D. H. Sharp, {\it Phys. Rev. Lett.} {\bf 58}, 
2162 (1987). 

\ref {29.}
T. H. Hansson, J. M. Leinaas and J. Myrheim, {\it Nucl. Phys. B} {\bf 384},
559 (1992).

\ref {30.}
D. Bonatsos and C. Daskaloyannis, {\it Phys. Lett. B} {\bf 278}, 1 (1992). 

\ref {31.}
D. Bonatsos and C. Daskaloyannis, {\it Chem. Phys. Lett.} {\bf 203}, 150 
(1993). 

\ref {32.}
D. Bonatsos, {\it J. Phys. A} {\bf 25}, L101 (1992). 

\ref {33.}
A. Klein and E. R. Marshalek, {\it Rev. Mod. Phys.} {\bf 63}, 375 (1991). 

\ref {34.}
V. Penna, {\it Mod. Phys. Lett. B} {\bf 5}, 1947 (1991). 

\ref {35.}
V. Barone, V. Penna and P. Sodano, {\it Phys. Lett. A} {\bf 161}, 41 (1991). 

\ref {36.}
V. Penna, {\it Phys. Lett. A} {\bf 125}, 385 (1987). 

\ref {37.}
V. Penna, {\it Physica A} {\bf 152}, 400 (1988). 

\ref {38.}
M. Rasetti and T. Regge, {\it Physica A} {\bf 80}, 217 (1975). 

\ref {39.}
M. Rasetti and T. Regge, {\it Quantum Vortices}, in {\it Highlights of 
Condensed Matter Theory}, ed. F. Bassani {\it et al.} (Compositori, Bologna, 
1985). 

\newpage
%\newcounter{tabn}
%\def\ds{\displaystyle}
%\def\llabel#1{\footnote{#1}\label{#1}}
%\def\llabel#1{\label{#1}}

%================begin table =======================================
\begin{table}[bth]
\begin{center}
\caption{ Structure functions of special deformation schemes}
\bigskip
\begin{tabular}{|c c p{2.0 in}|}
\hline
\ & $\Phi(x)$ & Reference \\
\hline\hline
\romannumeral 1 & $x$ & harmonic  oscillator \\[0.05in]
\romannumeral 2&  ${ {q^x- q^{-x} }  \over {q- q^{-1} } }= [x]_q $ &
$q$-deformed harmonic oscillator $^{13-15}$ \\[0.05in]
\romannumeral 3& ${ {Q^x- 1 } \over {Q- 1 } } = [x]_Q $ &  $Q$-deformed 
oscillator $^{11,12}$ \\[0.05in]
\romannumeral 4&  ${ {q^x- p^{-x} }  \over {q- p^{-1} } }=[x]_{p,q} $ &
$(p,q)$-deformed or  2-parameter oscillator $^{20}$
\\[0.05in]
\romannumeral 5& $x(p+1-x)$ & parafermionic oscillator
 $^{21}$ \\[0.05in]
\romannumeral 6& $ { \sinh (\tau x) \sinh (\tau (p+1-x) )}\over
{ \sinh^2(\tau) } $ & $q$-deformed parafermionic oscillator
$^{22}$  \\[0.05in]
\romannumeral 7& $x\cos^2(\pi x/2) + (x+p-1)\sin^2(\pi x /2)  $&
parabosonic oscillator $^{21}$ \\[0.05in]
\romannumeral 8&
$\begin{array}{c}
\frac{\sinh(\tau x)}{\sinh(\tau)} 
\frac{\cosh(\tau (x+2N_0-1))}{\cosh(\tau)} \cos^2 (\pi x/2) +\\
+ \frac{\sinh(\tau (x+2N_0-1))}{\sinh(\tau)} 
\frac{\cosh(\tau x)}{\cosh(\tau)} \sin^2 (\pi x/2)
\end{array}$
 & $q$-deformed parabosonic oscillator
 $^{22}$ \\[0.05in]
\romannumeral 9 &
$\sin^2 {\pi x/2}$ & fermionic algebra $^{23}$ \\[0.05in]
\romannumeral 10 & $ q^{x-1} \sin^2 {\pi x/2}$ &
$q$-deformed fermionic algebra $^{24}$ \\[0.05in]
\romannumeral 11& $ x(2-x) $ & fermionic algebra $^{21}$ \\[0.05in]
\romannumeral 12& $ {\sin(\tau x)\sin(\tau(2-x))\over \sin^2(\tau)} $ & 
$q$-deformed fermionic algebra $^{22,25}$ \\[0.05in]
\romannumeral 13&
$\frac{1-(-q)^x}{1+q}$ & generalized $q$-deformed fermionic algebra
$^{26}$ \\[0.05in]
\romannumeral 14 & $ [x]_{p,q} + 2\nu [x]_{-p,q} $ & modified 
$(p,q)$-oscillator $^{27}$ \\[0.05in]
\romannumeral 15& $x^n$ & $^{19}$ \\[0.05in]
\romannumeral 16& ${ {sn(\tau x)} \over {sn(\tau )} }$ & $^{19}$ \\[0.05in]
\hline
\end{tabular}
%\caption{\sf Structure functions of special  deformation schemes}
\end{center}
\end{table}
%%======================== end of table ======================

\end{document}